\documentclass[12pt]{article}
\usepackage{amsmath}
\usepackage{graphicx}
\usepackage{enumerate}
\usepackage{natbib}
\usepackage{url} % not crucial - just used below for the URL 
\usepackage{mathrsfs}
\usepackage{multirow}%
\usepackage{booktabs}
\usepackage{amssymb}
\usepackage{hyperref}
\usepackage{color}
\usepackage[export]{adjustbox}

\usepackage{float}

\newcommand{\blind}{1}

\newcommand{\ib}{\mathbf{i}}

\newcommand{\Ab}{\mathbf{A}}

\newcommand{\Eb}{\mathbf{E}}

\newcommand{\Lb}{\mathbf{L}}

\newcommand{\Vb}{\mathbf{V}}

\newcommand{\Xb}{\mathbf{X}}

\DeclareMathOperator*{\argmin}{arg\,min}

\begin{document}

\def\spacingset#1{\renewcommand{\baselinestretch}%
{#1}\small\normalsize} \spacingset{1}

%%%%%%%%%%%%%%%%%%%%%%%%%%%%%%%%%%%%%%%%%%%%%%%%%%%%%%%%%%%%%%%%%%%%%%%%%%%%%%

\if1\blind
{
  \title{\bf Multi-faceted Neuroimaging Data Integration via Analysis of Subspaces}
  \author{Andrew Ackerman$^1$, Zhengwu Zhang$^1$, Jan Hannig$^1$, Jack Prothero$^2$, J.S. Marron$^1$ %\thanks{
    %The authors gratefully acknowledge \textit{please remember to list all relevant funding sources in the unblinded version}}
    \hspace{.2cm}\\
    $^1$Department of Statistics and Operations Research,\\ University of North Carolina at Chapel Hill\\
    $^2$National Institute of Standards and Technology,\\
      Gaithersburg, MD, USA; 
    }
  \maketitle
} \fi

\if0\blind
{
  \bigskip
  \bigskip
  \bigskip
  \begin{center}
    {\LARGE\bf Multi-faceted Neuroimaging Data Integration via Analysis of Subspaces}
\end{center}
  \medskip
} \fi

\bigskip
\begin{abstract}
Neuroimaging studies, such as the Human Connectome Project (HCP), often collect multi-faceted and multi-block data to study the complex human brain.  However, these  data are often analyzed in a pairwise fashion, which can hinder our understanding of how different brain-related measures interact with each other.  In this study, we comprehensively analyze the multi-block HCP data using the Data Integration via Analysis of Subspaces (DIVAS) method. We integrate structural and functional brain connectivity, substance use, cognition, and genetics in an exhaustive five-block analysis. This gives rise to the important finding that genetics is the single data modality most predictive of brain connectivity, outside of brain connectivity itself.  Nearly 14\% of the variation in functional connectivity (FC) and roughly 12\% of the variation in structural connectivity (SC) is attributed to shared spaces with genetics.  Moreover, investigations of shared space loadings provide interpretable associations between particular brain regions and drivers of variability, such as alcohol consumption in the substance-use data block.  Novel Jackstraw hypothesis tests are developed for the DIVAS framework to establish statistically significant loadings.  For example, in the (FC, SC, and Substance Use) shared space, these novel hypothesis tests highlight largely negative functional and structural connections suggesting the brain's role in physiological responses to increased substance use. Furthermore, our findings have been validated using a subset of genetically relevant siblings or twins not studied in the main analysis.

%We analyze the multi-block Human Connectome Project data set using Data Integration via Analysis of Subspaces.  Numerous analyses have been done on this data, but largely in pairwise fashion.  By contrast, we integrate human brain connectivity, environmental features, and genetics in a comprehensive five-block analysis.  This gives rise to the important finding that genetics is the single data modality most predictive of brain connectivity, outside of brain connectivity itself.  Nearly 14\% of the variation in functional connectivity and roughly 12\% of the variation in structural connectivity is attributed to shared spaces with genetics.  Moreover, investigations of shared space loadings provide interpretable associations between particular brain regions and drivers of variability, such as alcohol consumption in the substance-use data block.  Jackstraw hypothesis tests are leveraged and extended to the DIVAS framework to establish statistically significant loadings features.  In the FC-SC-Use shared space, these novel hypothesis tests highlight largely negative functional and structural connections suggesting the brain's role in physiological responses to substance use. 

%symmetric (about hemisphere) functional connections including the brain stem that suggest the brain's role in physiological responses to substance use.  

\end{abstract}

\noindent%
\vfill

\newpage
\spacingset{1.9} % DON'T change the spacing!
\section{Introduction}
\label{sec:intro}

%Various types of imaging, behavioral, and genetic data are often collected on a common set of subjects in contemporary neuroscience studies.  Data sets arising from studies such as the Human Connectome Project (HCP), Adolescent Brain Cognitive Development (ABCD), and UK Biobank are examples of such multi-block (multi-omic, multi-view or multifaceted) data.   

%This work will focus on the Human Connectome Project (HCP) \citep{HCP} data set which includes brain structural connectivity (SC) and functional connectivity (FC) collected and estimated through diffusion and functional MRI.  

The Human Connectome Project (HCP) \citep{HCP} is a seminal study aimed at advancing our understanding of the human brain, particularly from the connectomic perspective. This work analyzes various data blocks present in the HCP Young Adult (HCP-YA) study in a more comprehensive manner than previously achieved.  Specifically, our analysis contains five different data blocks, including brain structural connectivity (SC) and functional connectivity (FC), which are collected and estimated through diffusion and functional Magnetic Resonance Imaging (MRI). Additional information on subjects' cognitive performance, substance use habits, and genetic composition is also analyzed in this multifaceted data integration case study.  The HCP dataset also presents the distinct merit of including first-order family relatives (parents and their offspring and/or siblings).  Splitting the data along these first-order relations provides a natural comparison between original and validation data sets and allows us to corroborate our findings as more than mere spurious associations.  

Many multi-block analyses of the HCP-YA data set have been informative in pairwise settings.  For example, \cite{intro1} aims to predict FC given SC using a 
higher order depedence measure.  \cite{ZHANG2022} uses multi-layer graph convolutional networks (GCN) within a generative adversarial network (GAN) to predict SC from FC.  \cite{finger} predicts cognition using functional connectivity, and \cite{gene_connect} links the human connectome to genetic heritability.  While these methods yield useful insights, they are restricted to consideration of only two modalities at a time -- a fact that limits our understanding of these likely interrelated data.  

The literature has, at times, ventured beyond this pairwise paradigm, as in the instance of \cite{intro3} investigating covariation between brain connectivity, demographic information (such as age, sex, and income), and behavioral traits (such as \textit{rule-breaking behavior}).  Moreover, \cite{HCPCCA} connects multiple types of brain connectivity with cognitive performance via Canonical Correlation Analysis (CCA) \citep{CCA}.  Likewise, \cite{cjive} integrates FC, SC, and fluid intelligence.  However, even in these more expansive analyses, consideration of either substance-use habits or genetic predispositions is absent. In this work, we extensively analyze the interrelation of FC, SC, cognition, substance use, and genetics using a state-of-the-art integration technique named Data Integration via Analysis of Subspace (DIVAS) \citep{divas}.

DIVAS uses a search through shared subspaces based on angle perturbation bounds, to distinguish signal from noise and further differentiate shared from partially shared and individual variation.  Accordingly, each data block included in the analysis is represented as a summation of low-rank matrices comprised of products of loadings and scores inherent to each signal subspace.  Specifically, we apply DIVAS to find fully shared, partially shared, and fully individual subspaces among the five HCP data blocks.  We also extended the Jackstraw Significance Testing from \cite{jackstraw} to identify statistically significant features within DIVAS segmented loadings.  Collectively, this yields biologically interpretable results while also highlighting the type of statistical inference that pairing these two methods (DIVAS and Jackstraw) can produce. 

The primary contributions of this work can be summarized as follows:

\begin{itemize}

\item Comprehensive analysis of relative signal strength corresponding to each data block.  Previous work has attempted to predict variation in cognition based on brain connectivity \cite{EFCSC}, or even predict SC given FC \cite{ZHANG2022} to understand how different data blocks or features are related with each other.  That said, being able to provide a specific percentage of signal strength available in \textit{each} data modality, FC through genetics, attributable to a particular segmented shared space represents a substantial advancement to the neuroscience literature. 

\item Confirmatory brain connectivity analysis with novel genetics and substance-use insights. Section \ref{varres} shows FC to be the most significant predictor of SC and vice versa \citep{ZHANG2022,intro1}.  Section \ref{varres} also depicts genetics as the second most influential data modality in determining brain connectivity, a result not previously established.  

%\item confirmatory analysis of brain connectivity with novel extensions.  Section \ref{varres} shows FC to be the most significant predictor of SC \cite{FCSC}.  The extent to which SC is predictive of FC is previously underdeveloped and is also shown in Section \ref{varres}.  Moreover, we can corroborate established findings at the brain region level.  For example, Section \ref{FCSCuse} will verify a link between the insula and substance-use \cite{insula}, while also suggesting this association is not only present in tobacco use (established and reproduced) but also alcohol dependence.  

%\item novel genetics and substance-use analysis. Section \ref{varres} depicts genetics as the second most influential data modality in determining FC,SC and substance-use.  Similarly, consistent with the above bullet point, Section \ref{FCSCuse} extends established results to a broader class of substance-use patterns.  

\item Extension of Jackstraw methodology to test statistical significance in DIVAS loadings. DIVAS loadings provide important insights on how different data blocks can vary with each other. The previous Jackstraw methodology defined in the AJIVE setting \citep{ajive} cannot be directly applied to DIVAS. Section \ref{js} will illustrate this new Jackstraw for the DIVAS framework. 

\item Results validation based on a separate HCP subset data.  The presence of first-order relatives in the HCP allows for a validation data set that is approximately an independent copy of the main discovery data set.  We then apply principal angle analysis to quantify the extent to which these subspaces, in potentially high dimensions, are reproducible.  Indeed, Section \ref{val} demonstrates that the results corresponding to the two data sets are highly related and that the subspaces discerned in the discovery set are reproduced by the validation set.

\end{itemize}

The remainder of the paper will be structured as follows: Section \ref{data} will discuss the data and associated preprocessing.  Section \ref{sec:meth} articulates the methods which entail DIVAS, Jackstraw, a variational decomposition, and principal angle analysis.   Section \ref{sec:Results} illustrates the results of applying these methods to the five-block HCP data, and Section \ref{sec:disc} concludes with discussion of our contributions and future work.  Additional diagnostic plots and discussion of bounds used in quanitifying reproducibility can be found in Appendix \ref{sec:app}.

%The confirmatory analysis will not only reaffirm established results but also provide credence for novel extensions.  
%applicability and validity of our chosen methods.  The novel analysis aims to more substantively contribute previously unestablished insights to the existing body of literature.  Finally, the validation on a paired subject (a member of the same first-order relation) demonstrates that while there is some level of person-by-person variation in HCP data, our results are largely reproducible and persistent across that variation.   

\section{Data}\label{data}

The Human Connectome Project Young Adult (HCP-YA)\citep{HCP} is a comprehensive neuroscientific study that has generated complex datasets on brain function, structure, cognitive performance, and more, involving more than 1,200 human subjects.  These data are freely accessible through the \href{https://www.humanconnectome.org/study/hcp-young-adult/data-releases}{ConnectomeDB website}.  The HCP is both expansive and highly structured in the sense that it contains first-order family relatives. Application of DIVAS to this HCP data allows for a more comprehensive analysis than has previously been accomplished, while also enabling a more related validation than is available in random partition methods.   

We preprocess five blocks of HCP data before applying DIVAS: SC, FC, substance use, cognition, and genetic measures.   In the subsections that follow, we explicitly detail what this preprocessing entails. 

\subsection{Preprocessing FC/SC Data}\label{prefcsc}

For each HCP-YA subject, we download dMRI, T1, and resting state fMRI (rs-fMRI) data. The dMRI session includes 6 runs, using 3 gradient tables (b=1000, 2000, and 3000), each acquired with opposite phase encoding polarities. Each table has approximately 90 diffusion-weighted directions and 6 interspersed b0. The scans were performed using a spin echo EPI sequence on a 3T Connectome Scanner, resulting in an isotropic voxel size of 1.25 $mm^3$ and 270 diffusion-weighted scans. The T1 image has 0.7 $mm^3$ isotropic resolution. See \cite{VANESSEN} for detailed acquisition and preprocessing information. We apply the population-based structural connectome mapping (PSC) framework \citep{method1} to the minimally preprocessed dMRI and T1 data to extract SC. PSC employs a reproducible probabilistic tractography algorithm \citep{maier2016tractography}, leveraging anatomical information from the T1 image to reduce tractography bias. We use the Desikan-Killiany (DK) atlas \citep{Desikan2006968} to define 68 cortical parcels, and the FreeSurfer template \citep{fischl2002whole} to define 19 subcortical regions, making a total of 87 regions of interest (ROIs). Streamlines connecting ROI pairs are extracted by dilating gray matter ROIs, isolating pathways by cutting streamlines, and removing outliers. Connectivity strength is quantified by the number of streamlines, a measure widely used in brain imaging-genetic studies \citep{chiang2011genetics, method2}.

The HCP-YA rs-fMRI data include two left-right and two right-left phase-encoded 15-minute eyes-open rs-fMRI runs \citep{VANESSEN}. Each run used 2 $mm^3$ isotropic voxels with a 0.72s repetition time. For each run, we calculate the average time series for each of the 68 cortical ROIs from \cite{Desikan2006968}, along with the 19 subcortical ROIs.  Pearson correlations between pairs of ROI's are computed for each run, Fisher z-transformed, averaged across the four runs, and transformed back to correlations.

DIVAS expects each data block to be a matrix $p_k \times n$ with $n$ human subjects in the columns and $p_k$ features along the rows, where $p_k$ is the number of features for the $k$-th data block.  Therefore, each connectivity adjacency matrix is vectorized before they can be stacked in its data block.  More specifically, the upper-triangular sub-matrix of each individual's symmetric adjacency matrix (both structural and functional) is vectorized and stacked horizontally to produce the columns of each connectivity data block.  Once created, this data block (matrix) is then \textit{object-mean} centered (\cite{ooda}) and variance thresholded.  Specifically, this centering entails subtracting the column vector whose entries are the means of the entries in the corresponding rows of the data matrix. 
%haven't introduced notation yet, so trying not to use ooda notation either 

%Importantly, each step of this adjacency matrix manipulation can be inverted so as to arrive at an adjacency matrix when beginning with a loadings vector.  This will be quite useful for visualization and interpretation in Section \ref{FCSCuse}.  Of the 1400 original imaging subjects, 1065 
%from 457 unique families 
%contain sufficiently complete imaging data to be included in DIVAS analysis. 

\subsection{Preprocessing Cognition/Substance-Use Data}

Cognitive performance measures are collected according to the cognition battery of test in the NIH Toolbox (\cite{nih}).  Ultimately, this data block contains 45 tests of cognitive performance from reading comprehension to spatial awareness collected across 1206 subjects.  Similarly, the substance-use data block contains 36 self-reported features ranging from frequency of alcohol use to age of first tobacco use.  Both the cognition and substance-use data blocks are object-mean centered, as described in Section \ref{prefcsc}, and normalized to further ensure that the scale of any one cognitive test or substance-use measure is not dominating DIVAS \textit{modes of variation}.
A mode of variation is a rank 1 matrix formed from the outer product of two vectors -- one in object space and one in trait space -- and will be extensively used in Section \ref{sec:meth}.  

Missing data are encountered in both blocks but most severely in the self-reported substance-use measures.  Any feature missing greater than half of its corresponding observations is removed.  This results in six tobacco and marijuana-use features  being removed, leaving 30 total substance-use features. No features are removed from the cognition data block.  The remaining missing data, in both cognition and substance-use, are filled using a simple row-mean imputation.  

\subsection{Preprocessing Genetic Data}

HCP participants provide blood samples from which a cell line can be created \citep{VANESSEN}.  Single Nucleotide Polymorphisms (SNPs) are extracted from these cell lines and made available on the database of Genotypes and Phenotypes (dbGaP) \footnote{\href{https://www.ncbi.nlm.nih.gov/gap/}{https://www.ncbi.nlm.nih.gov/gap/}} for each of 1141 subjects.  We have the preprocessed SNP data using methods from \cite{method2}.  
Specifically, any subjects missing more than 10\%
of its SNPs are removed from consideration. Additionally, any SNPs containing more than
5\% missing values, less than 5\% minor allele frequency, and a Hardy-Weinberg equilibrium
p-value less than $1x10^{-6}$ are excluded. The remaining data are further pruned using a linkage
disequilibrium-based method resulting in 130,452 SNPs. This is still a prohibitively large data block for DIVAS. Therefore, we apply principal component analysis (PCA) to the SNP data to extract the
first $d=n$ principal components as the final features.  This procedure entails a mere rotation of our data, so no data loss is involved. As such, the genetic data block consists of features that represent linear combinations of the already preprocessed SNPs and should be interpreted with due care.

%and the top $d=n$ principal components were retained as our ultimate features.  As such, our data block consists of features that represent linear combinations of the already significantly preprocessed
%SNPs and should be interpreted with due care. 

%Collectively, this preprocessing results in five data blocks of size $3591 \times 375$, $3509 \times 375$, $45 \times 375$, $36 \times 375$, and $375 \times 375$ for FC, SC, cognition, substance-use, and genetic SNPs respectively.  

\subsection{Unifying on Common (Non-related) Data Objects}

As Section \ref{divas} will discuss, DIVAS requires that the data blocks be unified on a common set of data objects -- in this case human subjects.  Practically, since each of the five data blocks above was collected on slightly distinct sets of subjects, a final required step is taking the set intersection of each subject list.  This winnows down the original 1206 subjects to 1064 common to all five data blocks.  

However, as discussed in Section \ref{sec:intro}, it is quite pivotal to note that the HCP data includes a large number of first-order family relations.  This poses serious challenges for any method, like DIVAS, that makes use of an independent observation assumption.  For that reason, we further reduce our sample by randomly selecting one representative from each unique family ID to arrive at 375 non-related individuals upon whom the independence assumption can more justifiably be applied.  This means that the finalized dimensions of the FC, SC, cognition, substance-use, and genetic data blocks are as follows: $3591 \times 375$, $3509 \times 375$, $45 \times 375$, $36 \times 375$, and $375 \times 375$. 

The validation data set analyzed in Section \ref{val} will be comprised of the same five data blocks, preprocessed in exactly the same manner described above but on a set of non-genetically related individuals.  The validation group will, however, be highly genetically related to the original group.  This is accomplished by taking a second random representative from each family ID, and ensuring that this representative indeed has collected information among all five data blocks.  
This results in a validation set of 377 individuals of which 326 are first order relatives with a member from the original set.

\section{Methods}\label{sec:meth}

We introduce methodology used to analyze the HCP data. DIVAS is implemented to integrate the five disparate data blocks, and is discussed in Section \ref{divas}.  Novel DIVAS Jackstraw Significance Tests are derived to assess the statistical significance of DIVAS loadings entries, and is discussed in Section \ref{js}.  A variational decomposition is used to describe the relative signal strength of each data block, and is discussed in Section \ref{varmeth}.  Finally, principal angle analysis is articulated in Section \ref{paa} as a method for assessing reproducibility.

\subsection{DIVAS}\label{divas}

 %DIVAS (\cite{divas}) finds shared (joint), partially shared, and individual subspaces within disparate data blocks.  Distinctively, DIVAS aims to find collections of rank 1 \textit{joint} subspaces, where joint here simply means having common scores.  

DIVAS \citep{divas} finds subspaces of $\mathbb{R}^n$ that represent either fully shared (joint), partially shared, or individual structure.  Basis vectors determine rank 1 \textit{modes of variation} for each type of subspace.  In this context, joint is defined in terms of common scores.  Before examining the algorithm in more detail, let us consider the modeling assumptions. 

Consider the following data model for $p_k \times n$ dimensional data matrix $\Xb_k$,

\begin{equation}
\label{eq:datamodel}
\Xb_k = \Ab_k + \Eb_k,
\end{equation}
where each data block is assumed to be the sum of a low-rank signal matrix $\Ab_k$ and full-rank noise matrix $\Eb_k$.  Additionally, to reflect shared and partially shared structure across data blocks we assume each $\Ab_k$ can be decomposed as 

\begin{equation}
    \label{eq:signalmodel}
    \Ab_k = \sum_{\ib | k\in\ib} \Lb_{\ib,k}\Vb_{\ib}^\top, 
\end{equation}
where $\Lb_{\ib,k}$ is the $p_k \times r_i$ \textit{loadings} matrix corresponding to the $k^{th}$ data block, $\Vb_{\ib}$ is the $n \times r_i$ \textit{common normalized scores} matrix, and the index extends over a power set $\ib\in 2^{\{1,\ldots,K\}}$.  For example, the loadings matrix for the second data block, associated with partially shared structure between the second and third data blocks is denoted $\Lb_{\{2,3\}, 2}$.  Whereas the scores matrix for this partially shared space is common to each data block and thus denoted $\Vb_{\{2,3\}}$ with no dependence on $k$.
We also denote the partially shared joint signal $\Ab_{\ib,k}=\Lb_{\ib,k}\Vb_{\ib}^\top$.
For a set of signal matrices $\Ab_1,...,\Ab_k$, Theorem 1 of \cite{divas} shows the existence and uniqueness of such a decomposition under mild conditions in Condition 1. 

With this model in place, let us more carefully consider the DIVAS algorithm. Broadly, DIVAS consists of three steps -- signal extraction, joint subspace estimation, and signal reconstruction.
The signal extraction step will employ random matrix theory and singular value decomposition to extract the magnitude of the signal as well as angle perturbation theory to establish its direction. Novel angle bounds are derived and estimated through a subspace rotation bootstrapping procedure. Collectively, this produces a low-rank approximation of the data matrix.  Crucially, this initial step is done on each data matrix separately but in both object ($\mathbb{R}^n$) and trait ($\mathbb{R}^{p_k}$) spaces.  This represents a unique inferential advancement as compared to precursor methods such as Angle-Based Joint and Individual Variation Explained (AJIVE) \citep{ajive}.

These estimated signal subspaces determine the objective function and constraints of a convex-concave optimization problem aimed at minimizing angular distance between candidate directions and subspaces.  In this step also, the inclusion of object space information is unique to DIVAS and allows for a heightened level of interpretability in the resulting shared space loadings vectors.  The precise formulation of the optimization objective and constraints can be found in \cite{divas} equation 2.8.  

Finally, each candidate direction is passed into step three which aims to reconstruct the signal matrices for each block. 
 This is accomplished by first concatenating all joint structure basis matrices induced by block $k$.  This concatenated basis matrix is then used in a linear regression to find the loadings for block $k$.  This precise linear regression is aimed at accounting for collinearity between partially shared spaces, and will be pivotal to Equation \ref{eq:js2} in Section \ref{js}.  Additionally, this step performs one final SVD projection along a direction of maximal variation.  This can be thought of as a re-rotation aimed at sorting the rank 1 modes of variation in order of importance.

\subsection{Jackstraw}\label{js}

A useful technique for understanding statistical significance of features in high dimensions is Jackstraw Significance Testing \citep{og_jack}.  This proposed hypothesis tests on the row-spaces of genomic loadings resulting from PCA.  \cite{jackstraw} extended the Jackstraw approach to the AJIVE setting \citep{ajive}.  Both of these types of inference are done on individual modes of variation which is not well suited for a subspace-based method such as DIVAS.

More specifically, when DIVAS estimates loadings, it needs to account for potential collinearity induced by partially shared spaces of the same block collection.  To do this, DIVAS, and by extension DIVAS Jackstraw, does not estimate loadings one individual mode of variation at a time but simultaneously. 
\cite{divas} denoted the estimated orthonormal basis (i.e. scores vectors) for the joint structure among blocks in collection $\ib$ as $\mathfrak{V}_{\ib}$.  For a given data block $k$, horizontally concatenate all joint structure basis matrices found involving block $k$ into one matrix $\left[\mathfrak{V}_{\ib}\right]_{\ib|k\in \ib} := \mathfrak{V}_{k}$.  Then $\mathfrak{L}_k$ is found by solving the following least square problem: 
 %$[\mathfrak{L}_{\ib,k}]_{\ib|k\in\ib}$ is chosen as the least-squares solution of the regression problem  
%In particular, we estimate $\mathfrak{L}_{\ib,k}$, the $p_k \times \sum_{i \in \ib}{r_i}$ dimensional \textit{concatenation} of all loadings matrices corresponding to block collection $\ib$, and $\mathfrak{V}_{\ib}$, the corresponding concatenation of scores matrices \citep{divas}.  Then the loadings can be simultaneously estimated via  
\begin{equation}
    \label{eq:js2}
    \mathfrak{L}_k = \argmin_{\mathfrak{L}}\|\Xb_k - \mathfrak{L}\cdot \mathfrak{V}_{k}^T||_2^2.
\end{equation} 
The columns of matrix $\mathfrak{L}_k$ can then be partitioned into loadings $[\mathfrak{L}_{\ib,k}]_{\ib|k\in\ib}$ corresponding to the columns of the score matrix  $\left[\mathfrak{V}_{\ib}\right]_{\ib|k\in \ib}$.  

Let $\mathfrak{L}\in \mathbb{R}^{p_k \times d}$ be a sub-matrix of $\mathfrak{L}_k$, whose columns represent a collection of modes of variation of interest. Typically, this would be either a single mode of variation or modes of variation corresponding to the entire data block $\mathfrak{L}_{\ib,k}$.  The former will be the specific formulation applied to attain the results in Section \ref{FCSCuse} 

We can then test whether the $i^{th}$ feature plays a role across any of the $d$ loading values of the matrix of interest $\mathfrak L$:
\begin{equation}\label{eq:Hypotheses}
H_0:  \mathfrak{L}_{i,j} = 0 \mbox{ for all } j \in \{1,...,d\}\quad\mbox{vs.}\quad
H_A:  \mathfrak{L}_{i,j} \neq 0 \mbox{ for at least one } j \in \{1,...,d\}.
\end{equation}
This is accomplished via an empirical F-test.  
At a high level, we calculate sum of squared differences between the observed response and the predicted response in \ref{eq:js2}, both with and without the modes of variation of interest.  Towards that end, 
define $\mathcal{S} = \sum_{i \in \ib}{\hat{r}_i}$, and let $\hat{\Xb}_{k}^1 = \hat{\mathfrak{L}}\hat{\mathfrak{V}_{k}^T}$ and $\hat{\Xb}_{k}^0 = \hat{\mathfrak{L}^0}(\hat{\mathfrak{V}}_{k}^0)^T$.  Here, $(\hat{\mathfrak{V}}_{k}^0)$ is the matrix $\hat{\mathfrak{V}}_{k}$ with the columns of $\mathfrak{L}^0$ removed, and $\hat{\mathfrak{L}^0}$ is the solution to \eqref{eq:js2} with $\mathfrak{V}_{k}^T$ replaced by $(\mathfrak{V}_{k}^0)^T$. For a fixed $i$, the corresponding sum-of-squares becomes: \begin{center}
    $SSE_{1i} = \sum_{j=1}^n{(\Xb_{k_{[i,j]}}-\hat{\Xb}_{k_{[i,j]}}^1)^2};
    SSE_{0i} = \sum_{j=1}^n{(\Xb_{k_{[i,j]}}-\hat{\Xb}_{k_{[i,j]}}^0)^2}$
\end{center} where $\Xb_{k_{[i,j]}}$ is the $[i,j]^{th}$ element of the $k^{th}$ data matrix, $\Xb_k$.  
Clearly, the sum of squares $SSE_{0i}$ is computed under the null hypothesis \eqref{eq:Hypotheses}.
Finally, the associated test statistics are given by: 
\begin{equation}
    \label{eq:js4}
    F_{i} = \frac{(SSE_{0i}-SSE_{1i})/d}{SSE_{1i}/(n-\mathcal{S})}.
\end{equation}

%$\hat{\Xb}_{k}^0 = \hat{\mathfrak{L}}_{[:,-l]}\hat{\mathfrak{V}}_{k; [-l,:]}^T$ 

%where $\hat{\mathfrak{L}}_{[:,-l]}$ denotes estimated loadings less the $l^{th}$ estimated direction(s).  As such, \\
%$\hat{\mathfrak{L}}_{[:,-l]} \in \mathbb{R}^{(p_k \times (\mathcal{S} -d))}$ and $\hat{\mathfrak{V}}_{k; [-l,:]}^T \in \mathbb{R}^{((\mathcal{S} -d) \times n)}$.  

%intuitively, we drop calculate sum of squared difference between the observed response and the predicted response in \ref{eq:js2}, both with and without a particular mode of variation.   These sums of squares can be used to determine of the contribution of each particular feature in a given mode of variation is substantial enough to be considered statistically significant.  That is, it comprises an empirical F-test statistic 

Because of the complex structure of the DIVAS Jackstraw loadings, we would not expect \eqref{eq:js4} to follow an $F$ distribution.  Instead, we will simulate a permutation based null distribution against which we compare our empirical $F$-test statistic.  In particular, to generate a sample from the null distribution of the $F$ statistic, we randomly select a feature $i$, permute the corresponding row (feature) of the original data matrix $\Xb_k$, fit the loadings using the permuted data, and compute the corresponding test statistics.  This is repeated $s\gg p_k$ times. For large $p_k$ this choice of $s$ can be computationally expensive. Therefore, following \cite{jackstraw}, this permutation can be done for $m$ rows simultaneously to speed up computation, but often at the expense of accuracy. For the analysis presented in Section \ref{FCSCuse}, $m=1$, $s = 15000$.  

Similarly, in principle, simulating this null distribution should be based on a complete rerun of DIVAS after each permutation.  However, as argued in \cite{jackstraw}, this would be extremely computationally expensive.  Moreover, in high dimensional data (such as the HCP data presented here), permuting a small number of features will have a minimal impact on the common normalized scores output from DIVAS.  Therefore, we concur with \cite{jackstraw} in recommending that the original DIVAS common normalized scores be used for each permutation step.

%We generate a sample of such null $F$ statistics, by randomly permuting $m$ rows of the original data matrix $\Xb_k$, and computing the corresponding test statistics.  Choice of $m$ can be any integer between $\{1,...,p_k\}$, but we recommend $m = 1$ and permuting the corresponding $i^{th}$ row of $\Xb_k$. This is repeated $s$ times for a total of $s \times m$ test statistics generated under the null hypothesis.  In principle, simulating this null distribution should be based on a complete rerun of DIVAS after each permutation.  However, as argued in \cite{jackstraw}, this would be extremely computationally expensive.  Moreover, in high dimensional data (such as the HCP data presented here), permuting a small number of features will have a minimal impact on the common normalized scores output from DIVAS.  Therefore, we concur with \cite{jackstraw} in recommending that the original DIVAS common normalized scores be used for each permutation step.   

%Full details of this simulation and efficiency improvements can be found in \cite{jackstraw}

We reject the null hypothesis if our observed $F$ test statistic is larger than the \\ $(1-\alpha)$ percentile of the null distribution.  Since we desire a test, not for a fixed $i$ but all $i \in \{1,..., p_k\}$, a \cite{bon} correction, dividing by the number of features in the corresponding data block, is suggested and used to account for multiple testing.  It is worth noting that this adjustment is known to be conservative.  Indeed, as a consequence of the union bound, it is a level $\alpha$ test irrespective of the dependence between p-values.  However, we find that in our analysis of the HCP, this conservative approach yields results with statistical \textit{and} practical significance.  That is to say, this correction achieves biologically interpretable results.  

%do we include discussion of efficiency correction?

%Finally, we want to make explicit a note on computational efficiency.  The full algorithm given in \cite{jackstraw}, requires \textit{recomputing} the common normalized scores for each permutation of the data block, which will in turn require a new run of the full AJIVE algorithm for each permutation. Depending on the size of the datablocks involved, this could be a highly non-trivial computational expense. This expense is only magnified when considering DIVAS, which is tasked with segmenting not only shared and individual spaces but also partially shared spaces.  More specifically, each run of DIVAS on the five-block HCP data takes somewhere between 2 and 4 days to complete.  Running the full Jackstraw algorithm on this data would then require thousands of these multi-day runs.  This is computationally intractable.  For this reason,
%Yang suggests, and we adopt here, the approximation method of simply 

\subsection{Variational Decomposition}\label{varmeth}

We will proceed with a sum-of-squares-like decomposition of each original data block.  More specifically, DIVAS produces a low-rank matrix approximation of each component (fully shared, partially shared, and individual) of a given data block’s signal.  The squared Frobenius norm of each low-rank matrix can be thought of as a measure of the \textit{energy} or variability inherent to the original data block that is attributable to said component.  For example, we could study the percent of variation in FC that is explained by its pairwise shared space with SC.  

Part of our purpose in presenting these variational decompositions will be to juxtapose naturally comparable data blocks, such as FC with SC and cognition with substance-use.  To do this, we will rely on a notion of \textit{relative signal strength} which in turn requires that we introduce the notation of estimated partially-shared signal matrix $\hat\Ab_{\ib,k}=\mathfrak L_{\ib,k} \mathfrak V_{\ib}^\top$ and $\hat{\Ab}_{k} = \sum_{\ib | k\in\ib} \hat \Ab_{\ib,k}$. %and corresponding decomposition in terms of non-residual signal: $\|\tilde{\Xb}_k\| _F^2 = \sum_{i \in \ib}{}{\|\Ab_{k_i}\|}_F^2$.  
Thus, the resulting ratio that measures relative signal strength in the $k^{th}$ block that the $k_i^{th}$ shared-space (individual space) contributes is

\begin{equation}\label{r2}
\tilde{R}^2_{k,i} = \frac{\|\hat{\Ab}_{\ib,k}\|_F^2}{\|\hat{\Ab}_k\| _F^2}.
\end{equation}
The relative shared strength for each data block is presented in Tables \ref{table1}, \ref{table2}, and \ref{table3} of \\ Section \ref{varres}.

%However, it is well-known that these blocks are not commensurately noisy.  Specifically, SC is known to be much sparser and noisier than its FC counterpart.  Likewise, as a result of being self-reported scores as opposed to objectively assessed metrics, substance-use will be much noisier than cognition.   For this reason, we will present the variational decomposition in terms of \textit{non-residual signal}.  That is, our $R^2_{k_i}$ value is really just the percentage of non-residual signal in the $k^{th}$ block that the $k_i^{th}$ shared-space (individual space) explains

\subsection{Principal Angle Analysis}\label{paa}

In this section, we provide a method for verifying reproducibility of results (discussed in Section \ref{sec:Results}) via a post-processing principal angle analysis.  Computing the \textit{principal angles} between subspaces is an established way to quantify angular closeness.  Following \cite{results8}, if $\mathcal{M}, \mathcal{N}$ are subspaces of $\mathbb{R}^d$ such that $dim(\mathcal{M})= m \leq n = dim(\mathcal{N})$, the principal angles $0^\circ \leq \theta_1 \leq \theta_2 \leq ... \leq \theta_m \leq 90^\circ$ are defined to satisfy:
%\theta_1 = min\left(cos^{-1}\left(\frac{|\langle x, y \rangle|} {||x||||y||}\right)\right), (x,y) \in \mathcal{M} \times \mathcal{N};\\
\begin{equation}\label{pa}
\theta_i = \min\biggr\{\cos^{-1}\left(\frac{|\langle x, y \rangle|} {||x||||y||}\right) \biggr\rvert (x,y) \in \mathcal{M} \times \mathcal{N}, x \perp x_j, y \perp y_j\ \forall j \in \{1,...,i-1\}\biggr\} 
\end{equation}
where $x,y$ are the corresponding \textit{principal vectors}.  All else being equal, comparatively small principal angles indicate subspaces that are closer to each other than those producing large principle angles.  As discussed in Section 16.2.2 of \cite{ooda}, our intuition regarding interpretation of angles degrades in higher dimensions.  In particular, subspaces that are similar can exhibit apparently large principle angles.  

DIVAS accounts for this with the \textit{random direction bound} described in Section 2.1.2 of \cite{divas}.  Intuitively, this provides a stochastic lower bound on the angle between randomly related subspaces.  In particular, the random direction bound is a low percentile of a null distribution created by taking angles between a fixed $\hat{r}$-dimensional subspace and unit vectors chosen uniformly at random.  
As such, any principal angle exceeding this random direction bound is considered large.  

This principal angle analysis and comparison will be computed for each subspace segmented in both the original and validation runs. Any principal angle below the random direction bound gives indication of reproducibility, and any subspace with a majority of such principal angles shows rigorous evidence of overall reproducibility.

\section{Results}
\label{sec:Results}

\begin{figure}[h]
\begin{center}
\includegraphics[width=4in]{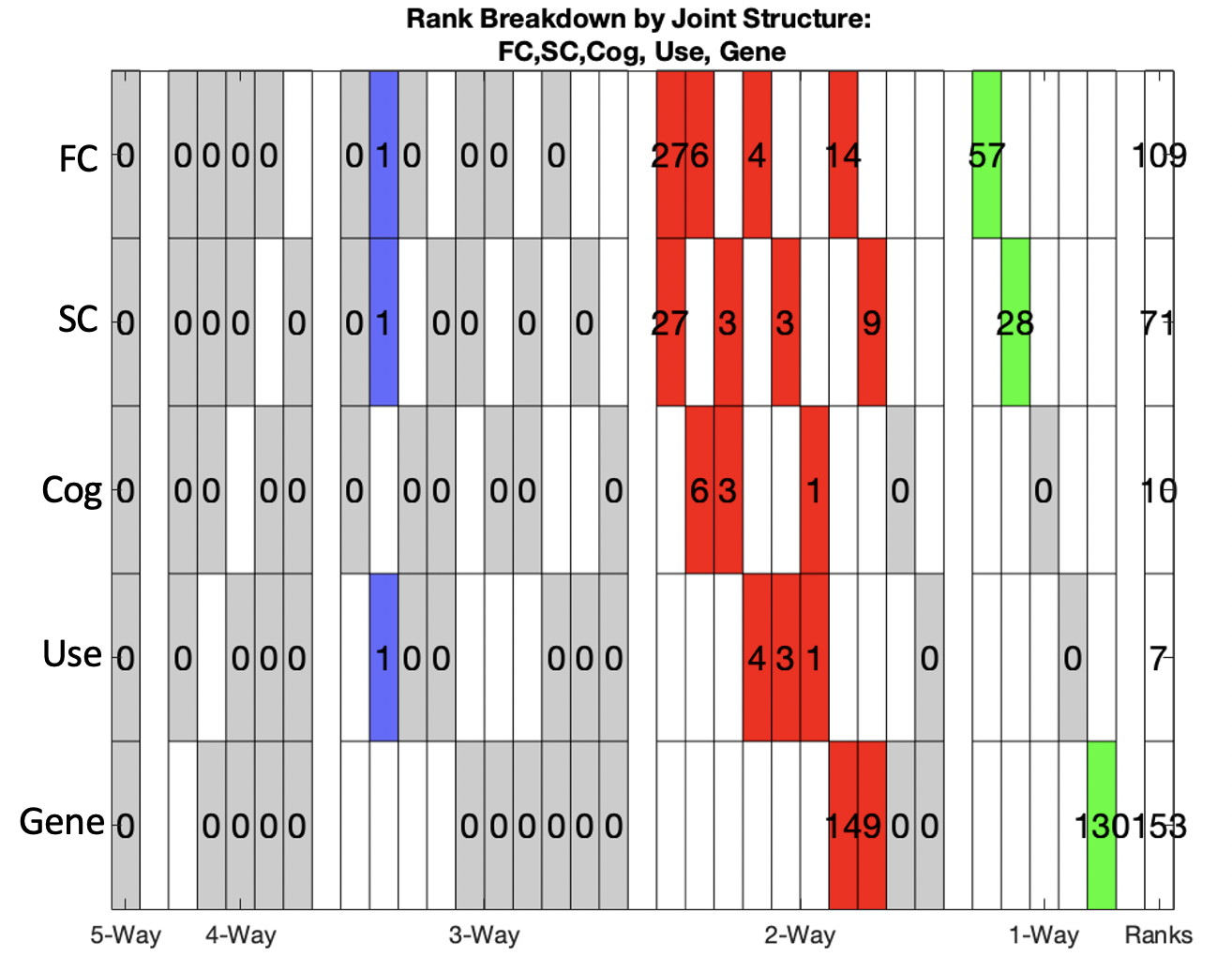}
\end{center}
\caption{DIVAS diagnostic plot for five-block run on FC, SC, Cognition, Substance-Use, and Genetics. \label{fig:diag}}
\end{figure}

The DIVAS and Jackstraw methods articulated in Sections \ref{divas}, \ref{js} are applied to the HCP preprocessed data as described in Section \ref{data}.  Figure \ref{fig:diag} illustrates a DIVAS diagnostic plot for this five-block run.
Each row represents a different data block, while each column represents a different type of shared, partially shared, or individual space.  The number within each cell represents the rank of the space segmented such that there is a rank 1 FC-SC-Use space, a rank 27 FC-SC space, etc.  Different colors are used to visually distinguish each type of segmented space, with a grey zero indicating a space that was indistinguishable from pure noise thus not segmented.  This diagnostic indicates no fully shared five-way or partially shared four-way spaces, one partially shared three-way space, a host of pairwise spaces, and three individual spaces.

The results section will proceed as follows: a variational decomposition aimed at describing how each shared-space contributes to explaining variability in a particular data modality, a careful interpretation of Jackstraw significant loadings in the FC-SC-Use subspace to elucidate biological interpretations of our HCP analysis, and a principal-angle validation routine verifying the robust nature of these findings.

\subsection{Variational Decomposition}\label{varres}
%lead into this now that method is above
\begin{table}[h]
\caption{FC/SC Shared Space Variational Decomposition}\label{table1}
\begin{tabular*}{\textwidth}{@{\extracolsep\fill}cccccc}
\toprule%
& \multicolumn{2}{@{}c@{}}{Functional Connectivity} & \multicolumn{2}{@{}c@{}}{Structural Connectivity} \\\cmidrule{1-3}\cmidrule{4-6}%
 Shared Space & $\tilde{R}^2$ & Rank & Shared Space & $\tilde{R}^2$ & Rank \\
\midrule
 Ind. FC & 51.14\% & 57 & SC-FC & 40.27\% & 27\\
 FC-SC & 23.45\%  & 27  & Ind. SC & 38.52\% & 28\\
 FC-Gene & 13.85\%  & 14  & SC-Gene & 11.58\% & 9\\
 FC-Cog & 7.72\%  & 6  & SC-Cog & 4.36\% & 3\\
 FC-Use & 3.16\%  & 4  & SC-Use & 3.87\% & 3\\
 FC-SC-Use & 0.67\%  & 1  & SC-FC-Use & 1.39\% & 1\\
\bottomrule
\end{tabular*}
\footnotetext{ }
\end{table}

Table \ref{table1} shows this decomposition applied to FC and SC.  Unsurprisingly, the single most influential shared space in FC and SC alike, is the pairwise space they share with each other.  Roughly 24\% of the variation in non-residual signal in FC can be attributed to a shared space with SC, while about 41\% of this variation in SC can be attributed to a shared space with FC.  Such significant portions of variation explained in each type of connectivity corroborates \cite{intro1} and \cite{ZHANG2022} in predicting FC with SC or vice versa.

Table \ref{table1} also illustrates the precise extent to which genetics contributes to understanding brain connectivity.  Genetics contributes the second most influential partially shared space in explaining both FC and SC.  This pairwise space has a relative signal strength of 13.85\% and 11.58\% in FC and SC respectively.  To the best of our knowledge, no previous work has established precise measures of the variability in brain connectivity attributable to genetic SNP's.  Genetics explaining such a non-trivial portion of variation indicates that both anatomical brain structure, such as white matter tracts, and the functional associations therein, are heavily influenced by genetic predisposition.

\begin{table}[h]
\caption{Cog/Use Shared Space Variational Decomposition}\label{table2}
\begin{tabular*}{\textwidth}{@{\extracolsep\fill}cccccc}
\toprule%
& \multicolumn{2}{@{}c@{}}{Cognition} & \multicolumn{2}{@{}c@{}}{Substance-Use} \\\cmidrule{1-3}\cmidrule{4-6}%
 Shared Space & $\tilde{R}^2$ & Rank & Shared Space & $\tilde{R}^2$ & Rank \\
\midrule
 Cog-FC & 62.30\% & 6 & Use-FC & 38.55\% & 4\\
 -- & --  & --  & Use-FC-SC & 34.25\% & 1\\
 Cog-SC & 31.74\%  & 3  & Use-SC & 19.55\% & 3\\
 Cog-Use & 5.96\%  & 1  & Cog-Use & 7.65\% & 1\\
\bottomrule
\end{tabular*}
\footnotetext{ }
\end{table}

Table \ref{table2} gives an analogous decomposition for the cognition and substance-use data blocks.  Of note, FC continues to be exceedingly important when explaining substance-use and cognition.  The pairwise partially space shared with FC is the single most informative space in determining both cognition and substance-use.  Moreover, between the pairwise Use-FC space and the three-way Use-FC-SC space, 72.80\% of the relative signal strength in substance-use is attributed to a partially shared space included functional connectivity.  Likewise, 62.30\% of the relative signal strength in cognition is attributed to a (pairwise) partially shared space with FC.  

Furthermore, structural connectivity has a non-trivial role to play in explaining cognition (31.74\%) and substance-use (19.55\% collectively).   This underscores the extent to which brain connectivity explains cognitive performance and substance-use patterns.   %citation.

\begin{table}[h]
\caption{Genetics Shared Space Variational Decomposition}
\label{table3}
\begin{tabular*}{\textwidth}{@{\extracolsep\fill}ccc}
\toprule
%& \multicolumn{3}{c}{Genetics} \\
\cmidrule{1-3}
Shared Space & $\tilde{R}^2$ & Rank \\
\midrule
Ind. Gene & 83.04\% & 130 \\
Gene-FC & 9.50\% & 14 \\
Gene-SC & 7.46\% & 9 \\
\bottomrule
\end{tabular*}
\end{table}

We conclude this variational decomposition section by applying equation \ref{r2} to the genetics data block, the results of which can be found in Table \ref{table3}.  Genetics, somewhat like cognition, is a data block whose signal was only segmented into comparatively few subspaces.  In particular it has an individual subspace and two pairwise subspaces.  Of these two pairwise partially shared spaces, FC accounts for the most variation in genetics, but  
brain connectivity as a whole contributes roughly 17\% of the non-residual signal variability in genetics.  Interestingly, no cognition or use shared-space was segmented, indicating that for this group of HCP subjects, genetics does not seem to explain cognition or use, except indirectly through brain connectivity.

\subsection{Investigation of Shared Spaces}\label{FCSCuse}

\begin{figure}[h]
\begin{center}
\includegraphics[width=4in]{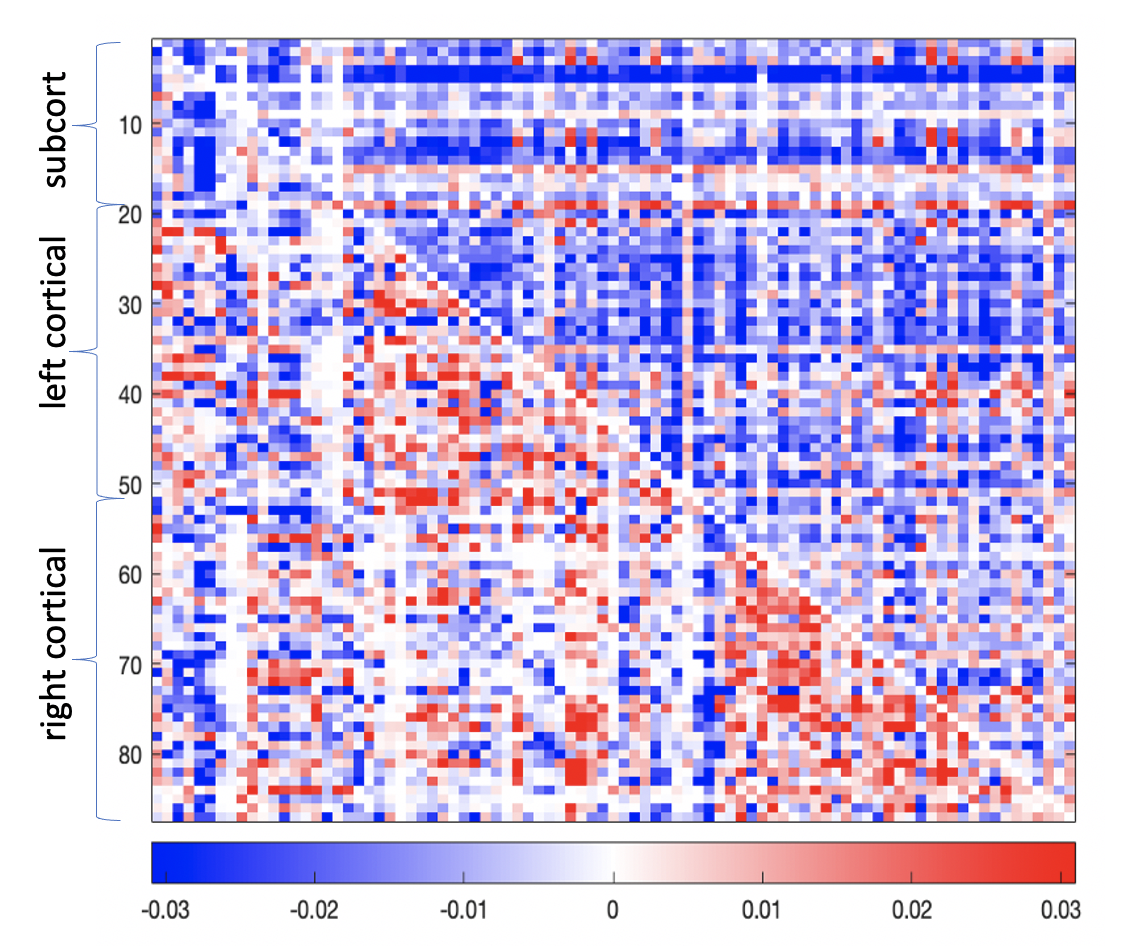}
\end{center}
\caption{FC and SC loadings adjacency matrix corresponding to the rank 1 FC-SC-Use subspace. The upper triangular represents the FC connections, and the lower triangular represents the SC connections.  Hence this matrix is not symmetric.  SC is more sparse than FC, but both FC and SC appear to be driven by predominantely negative connections. 
 }\label{fig:adj}
\end{figure}

%\begin{figure}[h]
%\begin{center}
%\includegraphics[width=3in]{fcscuse.png}
%\end{center}
%\caption{Note: we start to conflate jackstraw reduced and raw loadings showing these together.  Also, don't think it is worth it to show the raw bar chart and the reduced bar chart  
%}\label{fig:adj}
%\end{figure}

Investigating the loadings inherent to particular shared spaces allows for insight at the level of specific features rather than entire data modalities.  We begin by analyzing the rank 1 partially shared space between FC-SC-Use for two reasons.  First, it is the subspace containing the contribution from the most data modalities.  
In particular, it is the subspace with the most shared blocks (3).  
Secondly, while each segmented shared or individual space represents a statistically significant subspace, this subspace will be shown to be highly biologically interpretable as well. 
%practically significant as well.  
%In this case, practical significance comes in the form of biological interpretability.  Therefore, we will demonstrate that the FC-SC-Use subspace has a high level of such biological interpretability to offer.  

\begin{figure}[h]
\begin{center}
\includegraphics[width=6.0 in]{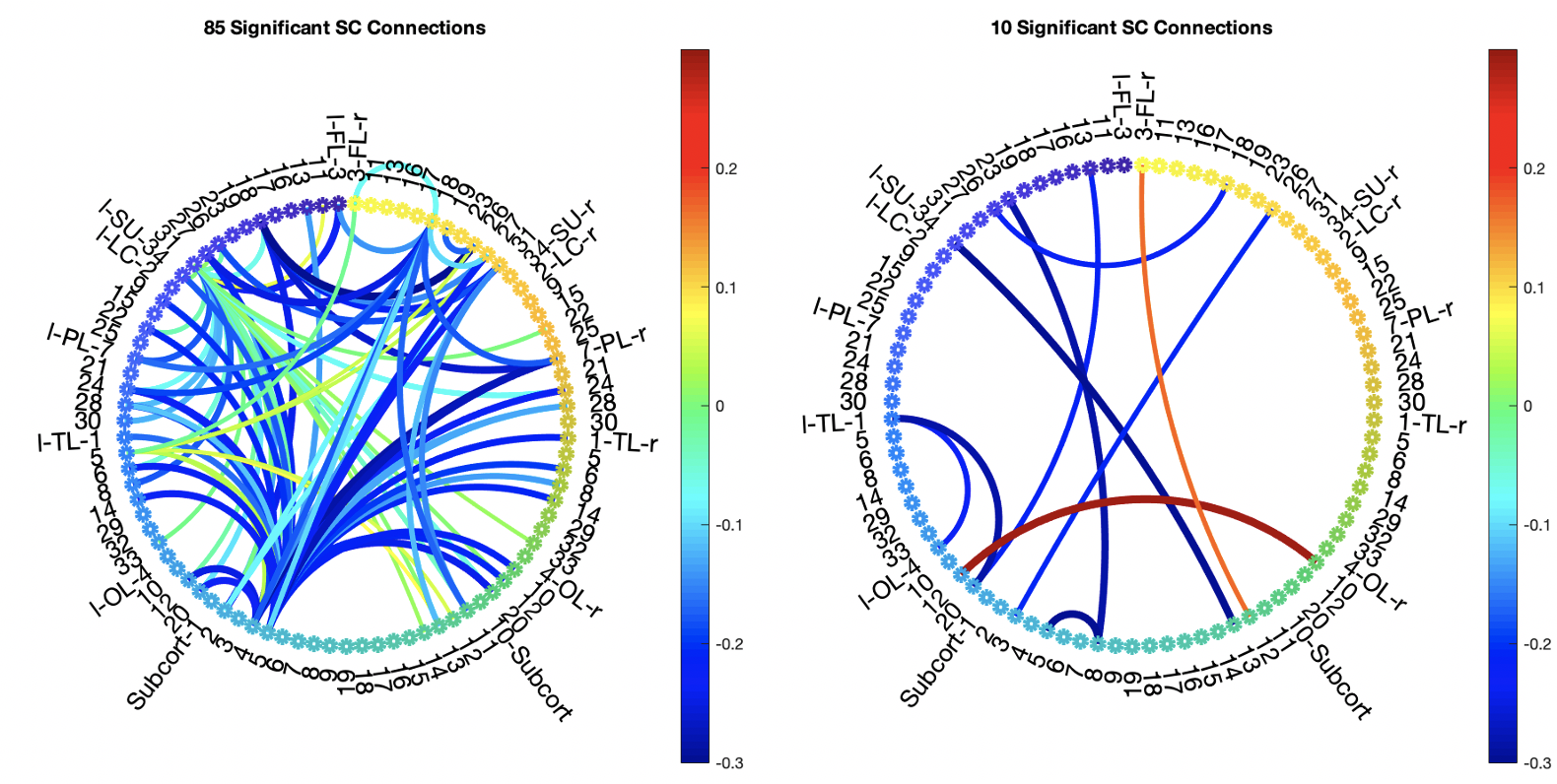}
\end{center}
\caption{FC (left) and SC (right) significant connections in rank 1 FC-SC-Use Subspace.  FC regions are reordered to correspond to SC regions. 
 These regions correspond to the adjacency matrix in Figure \ref{fig:adj}. 
 Abbreviations are used to denote brain regions as in FL (frontal lobe), PL(Parietal Lobe), OL (Occipital Lobe), and TL(Temporal Lobe). 
 Left and right hemispheres are denoted by 
``-l" and ``-r" respectively, and the subcortical regions are distinguished from the cortical regions by ``Subcort". 
 Observe that the vast majority of both FC and SC significant connections are negative.}\label{fig:circ}
\end{figure}

\begin{figure}[h]
\begin{center}
\includegraphics[width=5.5in]{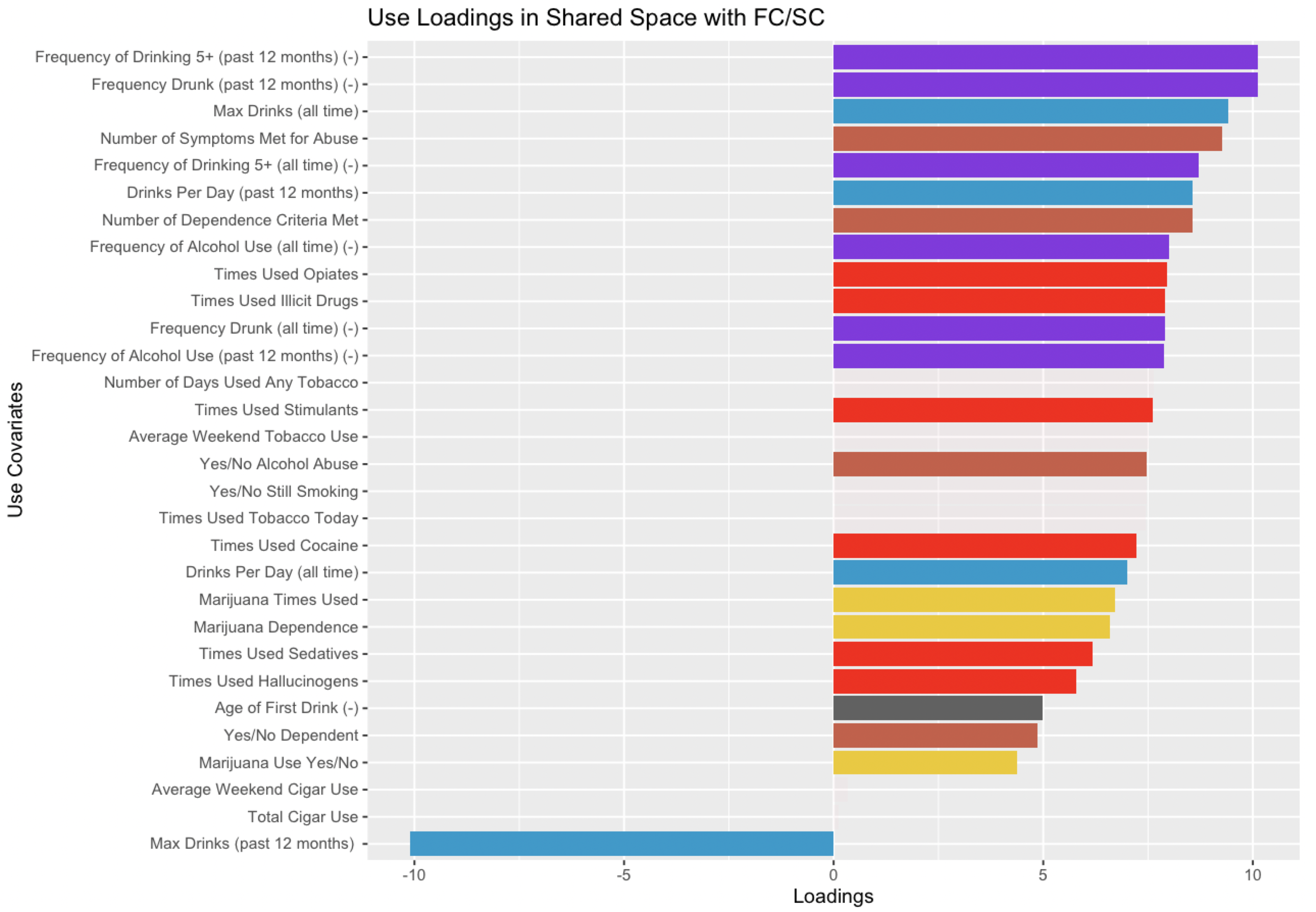}
\end{center}
\caption{Substance Use loadings corresponding to rank 1 FC-SC-Use partially shared space.  Jackstraw significant features will have full opacity while insignificant features are made translucent.}\label{fig:use}
\end{figure}

%Figure \ref{fig:load} shows the back-transformed FC loadings corresponding to the FC-SC-Use partially shared space.  Positive connections are denoted in shades of red, negative connections in shades of blue, and zero connections in white.  The 87 total brain regions are ordered to correspond to the SC regions presented in Figure \ref{fig:SC}.  It is worth noting that these loadings have not had Jackstraw applied to them yet.  Doing so will specify which of the 3591 connections are most worth considering in their relation to SC and substance-use.  
Figure \ref{fig:adj} shows the shows the back-transformed FC and SC loadings corresponding to the FC-SC-Use partially shared space.  This has been consolidated into a single adjacency matrix with FC connections on the upper triangular sub-matrix and SC connections on the lower.  Rows 1-9 represent left subcortical regions, while rows 10-18 represent right subcortical.  Row 19 represents the subcortical brain stem.  Similarly, rows 20-53 represent left cortical regions, and remaining rows 54-87 represent right cortical regions. There are several key observations to make from this adjacency matrix.  Firstly, SC is far more sparse than FC, which is consistent with established intuition.  Secondly, FC especially and SC to a lesser extent is dominated by negative (blue) connections.  It is important to note that Jackstraw Significance Tests have not yet been applied to these loadings, so if we compare Figure \ref{fig:adj} with Figure \ref{fig:circ}, we observe that the vast majority of both the FC and SC significant connections are indeed negative.  This signage will become quite important when we interpret it alongside the use loadings depicted in Figure \ref{fig:use}.  

Figure \ref{fig:circ} itself shows a circle plot of the 85 Jackstraw significant FC connections and 10 Jackstraw significant SC connections corresponding to the rank 1 FC-SC-Use partially shared space.  This may seem to be a minimal number of significant connections given the total number of features involved, however this is defensible for two reasons.  As stated in Section \ref{js}, a Bonferroni level $\alpha = 0.05$ adjustment is used to correct for multiple testing, and this adjustment is known to be conservative, meaning that it yields an underestimate of statistical significance.  %In this context, a conservative estimate is suitable insofar as claiming statistical significance where there is none seems more damaging to biological interpretation, than the opposite error. %-- claiming statistical insignificance when, in reality, the feature is significant.  
Consequently, we assert, and indeed the features deemed significant seem to confirm, that this procedure produces intuitive biological results.  

The most compelling interpretations will come from pairing these Jackstraw significant connections with the Jackstraw significant use features, but there are some high level observations to be gleaned from Figure \ref{fig:circ}.  As previously mentioned, both FC and SC significant features are dominated by negative connections.   Several of the largest in magnitude negative FC connections include subcortical region 5 -- the left putamen.  Both association (within hemisphere) and commissural (connecting left and right hemisphere) connections are present among these significant features.  Likewise, of the 10 significant SC connections 4 are commissural, and 6 are associative. Large connections between subcortical 5 (left putamen) and subcortical 8 (left amygdala) as well as between left 34 (insula) and subcortical 13 (right caudate) will be investigated, as they appear particularly influential.  

%\begin{figure}[h]
%\begin{center}
%\includegraphics[width=4in]{SC.png}
%\end{center}
%\caption{SC loadings corresponding to the FC-SC-Use rank 1 partially shared space.  Brain regions have been ordered to correspond with the FC regions shown in Figure \ref{fig:load}}\label{fig:SC}
%\end{figure}

%Figures \ref{fig:SC} and \ref{} show the adjacency matrix and circle plots that are SC analogues of Figures \ref{fig:load} and \ref{} respectively.  Jackstraw has again been applied on the adjacency matrix and is depicted in the circle plot to illustrate the significant connections at an $\alpha = 0.05$ level with Bonferroni adjustment. 

Figure \ref{fig:use} depicts the substance use loadings corresponding to the rank 1 FC-SC-Use partially shared space.  Jackstraw significant features are given full opacity while insignificant features are made translucent.  Moreover, features are color-coded according to type of substance use, and inversely coded %\footnote{By \textit{inversely coded} we mean a larger score in this feature indicates less, rather than more, substance use.} 
features are denoted by (-) and flipped to ease visual interpretation.  By inversely coded, we simply mean a larger score on such a features indicates less, rather than more, substance-use.  For example, \textit{Frequency of Drinking 5+} is a categorical variable, coded such that a 1 signifies the most frequent alcohol use and 7 signifies no use whatsoever.  It is, therefore, inversely coded and has had its sign flipped in Figure \ref{fig:use} to be consistent with the signage of conventionally coded features such as \textit{Number of Symptoms Met for Alcohol Abuse}. 

This substance use loadings is predominately driven by alcohol use features, and to a lesser extent marijuana and illicit substance use.  Also notice that the bar chart is largely positively oriented.  The combined interpretation of the signage in Figures \ref{fig:circ} and \ref{fig:use} is that an individual with a large score will exhibit more pronounced substance use as well as fewer blue connections and more red connections.  Therefore, the dominance of negative connections in FC and SC loadings lends the intuitive interpretation that substance use (alcohol in particular) is generally detrimental to brain connectivity.  We will explicitly examine the few connections that stand as exceptions to this finding.  

We also aim to provide integrated interpretations by linking individual connections to specific substance-use patterns. Figure \ref{fig:circ} illustrates a significant negative FC connection between subcortical region 5 (left putamen) and right cortical region 7 (right inferior parietal lobe).  Lessened functioning of the left putamen has been linked, through reward processing and motivation, to increased substance use \citep{putamen}.  Similarly, \cite{parietal} demonstrated that lessened activity in both the putamen and (bilateral) inferior parietal lobe are predictive of heightened substance use.  A second large and significant negative FC connection exists between subcortcial region 5 and left cortical region 24 (left precuneus).  Greater activation of the precuneus region has been shown to lessen the craving cues that are associated with alcohol \citep{precuneus1} and cannibis use \citep{precuneus2}.  

Turning to significant negative SC connections, subcortical region 13 (right caudate) is connected to cortical region 34 (left insula).  Lessened insular activity is associated with heightened risk for addiction \citep{revisit}.  Moreover alcohol dependence, specifically, has been linked to lessened functional activation in the caudate \citep{caudate_alc}.  Finally, structural connections between these regions have been shown to play a role in decision making and pain management \cite{SC_connect}.  Collectively, this provides strong evidence for our result that lessened structural connectivity between these two regions is associated with heightened substance use (particularly alcohol dependence) -- perhaps as a result of impaired pain management.  A second influential negative structural connection involves subcortical 5 (left putamen) and subcortical 8 (left amygdala).  Our analysis of functional connectivity also highlighted the impact of the left putamen, and the discussion in the preceding paragraph links this region to substance use \citep{putamen}, \citep{parietal}.  Similarly the amygdala, as a hub for reward processing, exhibits marked ``disregulation" in the wake of chronic substance use (including alcohol dependence) \citep{amy}.  Taken together, our results suggest that lessened structural connectivity between the amygdala and putamen is associated with heightened substance use.  In light of \cite{sc_put_amy} associating the structural connection between these two regions to pain management and memory of said pain, our results seem far from unfounded.

%Turning to significant negative SC connections, subcortical region 7 (left hippocampus) is connected to cortical region 24 (right precuneus).  The hippocampus has been linked to contextual memories which can trigger relapse in those with substance use disorders \citep{hippo}.  Moreover, a functional connection between the hippocampus and ventral striatum has been positively associated with substance use (alcohol and marijuana use in particular) in adolescence \citep{hip2}.  We present the related but distinct finding that lessened structural connectivity between the hippocampus and precuneus are associated with heightened use, indicating that perhaps this structural connection that relates to cue-induced cravings plays an inhibitory role.  A second influential connection involves subcortical region 11 (cerebellum-cortex) with cortical region 22 (right posterior cingulate).  Studies have associated deterioration of functional connectivity including the anterior cingulate cortex with increased substance use \citep{cing}.  Our results would suggest that lessened structural connectivity including the adjacent posterior cingulate cortex is also associated with heightened substance use.  Moreover, \cite{can} finds that the posterior cingulate contributes to cue-reactivity for cannabis use disorder.  \cite{cere} argues that the cerebellum-cortex has a role to play in addiction disorders through its role in reward processing and emotional regulation.  

Despite circle plots that are overwhelmingly negative, a minority of positive connections exist.  These connections indicate that heightened connectivity in the corresponding regions is associated with more pronounced substance use.  
Left region 12 (lingual) is positively connected to right region 10 (lateral occipital) -- both parts of the occipital lobe.  \cite{occipital1} has linked the occipital lobe, as a whole, to alcohol and cannabis use (independently).  However, these results were for a more general set of regions than we describe, and they report a ``blunted occipital alpha response".  Our results seem to inspire further investigation into the structural connectivity of these specific sub-regions of the occipital lobe.  Even so, it is conceivable that sensory perception is temporarily increased while under the effects of stimulants (also a significantly positive substance use covariate in Figure \ref{fig:use}).  If this were the case, a positive connection between such brain regions would be intuitive in light of a positive stimulant covariate.  Lastly, cortical region 3 (the right caudal middle frontal lobe) is positively connected with subcortical region 12 (right thalamus).  \cite{positive1} argues that \textit{increased} activity in the thalamus is observed in humans when reacting to drug cues and reduced during response inhibition.  Also, \cite{positive2} has shown the orbitofrontal cortex (of which the caudal frontal lobe is a part) is \text{activated} in addicted individuals during ``intoxication, craving, and bingeing" while being deactivated during withdrawal.  Consequently increased structural connectivity between two such regions could indeed be intuitively associated with heightened substance use.  

%Beginning with the SC circle plot in Figure \ref{fig:circ}, cortical region 5 (left entorhinal) is connected to cortical region 29 (left superior temporal lobe).  While possible interactions between the superior temporal lobe and substance use do not seem to have appeared in the literature, high concentration of cannabinoid receptors within the entorhinal region have been associated with heightened cannabis use \citep{ent}.  Consequently, increasing structural connectivity between a region known to harbor such receptors could indeed be intuitively associated with heightened substance use.  Likewise, FC connection 19 (brain stem) to 21 (post central gyrus) is notably positive.  The post central gyrus helps to regulate sensory awareness \citep{post}.  Degradation of the brain stem has also been shown to impair sensory integration \citep{bs}.  Therefore, we speculate that this significant positive connection could represent (temporary) increased sensory perception while under the influence of stimulants (another significant substance use feature) \citep{photo}, \citep{meth}.

In totality, our five-way analysis produces a rank 1 three-way shared space between FC-SC-Use with biologically interpretable results.  Specifically, we have found statistically significant negative connections that seem to corroborate the role of individual brain regions, or even connections more generally.  Similarly, the presence of the specific minority of positive connections observed is quite consistent with what is known about the role of the corresponding brain regions.  Even if the role of constituent brain regions were previously established, several of the \textit{connections} emphasized by our analysis have a previously unobserved association with substance use.

\subsection{Validation}\label{val} 

We present a principal angle analysis (Section \ref{paa}) comparing the original 5 block DIVAS run with the validation 5 block run, both described in Section \ref{data}.  Table \ref{table4} shows the principal angle analysis in each loadings space corresponding to a subspace that is shared across original and validation runs.  The corresponding \textit{minimum} principal angle between subspaces is listed in the third column, while the fourth column lists the fraction of principal angles in a given subspace that fall below the random direction bound (Section \ref{paa}). The more thorough DIVAS diagnostic plots for the original and validation runs are given in Appendix \ref{sec:app} and include the aforementioned random direction bound as a dot-dashed line in each cell.   

Of the 11 subspaces consistently segmented across original and validation runs, 9 exhibit a majority of associated principal angles falling below the random direction bound and therefore appear quite reproducible.  Brain connectivity loadings, collectively, represent $\frac{137}{156} \approx 88\%$ principal angles below the corresponding random direction bound.  Likewise, genetics loadings contain $\frac{69}{98} \approx 70\%$ principal angles below its random direction bound.  Finally, cognition loadings exhibit $\frac{7}{10} = 70\%$ principal angles below the random direction bound, and $\frac{4}{7} \approx 57\%$ of use loadings principal angles are less than their random direction bound.  This provides strong evidence of the general reproducibility of our analysis, both at the subspace and loadings level.  More specifically though, of the two subspaces that do not appear reproducible, the pairwise Cog-Use subspace warrants more discussion.   This is because it is the only consistently segmented subspace where none of its loadings directions fall below than random direction bound, and because its lack of reproducibility seems to have a telling explanation.  Consequently, we will restrict our attention to the Cog-Use pairwise space.  

Figures \ref{fig:diag_full} and \ref{fig:valid_full} illustrate that while the original and validation runs are remarkably similar in shared spaces segmented, the single three-way partially shared space segmented in the original run was FC-SC-Use while in the validation run it was FC-SC-Cog.  Moreover, when further investigating the principal angles between the connectivity loadings involved in these shared spaces, the FC components exhibit principal angles that fall well below the random direction bound.  Thus, it would appear that the functional connectivity portion of these subspaces are reproducible, but there persists some interaction between connectivity and use that is not replicated in the validation run (which in turn, exhibits some interaction between connectivity and cognition).  This has bearing on the pairwise Cog-Use subspace because DIVAS segments higher-order spaces first.  Specifically, the three-way subspaces are segmented prior to the pairwise subspaces, and the pairwise subspaces aim to account for variation that is left unexplained by the three-way (or higher) subspaces.  Therefore, when the three-way spaces exhibit slightly different interactions across use and cognition, it only stands to reason that the cognition and use pairwise spaces are going to have different left-over variation to explain.  

In totality, the large amount of confirmatory analysis produced in Section \ref{FCSCuse}, alongside the overwhelming majority of principal angles in Table \ref{paatable} indicate the reproducibility of our results.  The principal angle analysis, specifically, is particularly rigorous mechanism for assessing reproducibility.  Our models' performance with respect to this metric underscores the unusual precision of our analysis.  Future work is warranted in understanding what sorts of interactions persists between cognition, brain connectivity, and substance use, but the presence of such interactions should not significantly hamper the credence of our findings.  

\section{Discussion}\label{sec:disc}

This work represents numerous advancements to the neuroscience and statistics literature.  At the most general level, no previous HCP investigation has been as comprehensive as the above 5 block analysis.  Within it,  we corroborate established results such as the predictive capacity of FC on SC (\cite{ZHANG2022}), %and the role that substance-use plays in hampering social cognition (\cite{empathy1}), 
while also also establishing original results including the precise role genetics plays in predicting whole brain connectivity (Section \ref{varres}).  % and the role of a functional connection between the cuneus and brainstem in regulating pupilary response to alcohol use (Section \ref{FCSCuse}).    
Moreover, the methodology presents several unique insights.  Our Jackstraw methodology is a substantial abstraction from existing methods \citep{jackstraw} to take full advantage of the rich structure of DIVAS loadings.  Similarly, the variational decomposition 
uses non-residual signal as an elegant measure of relative signal strength across disparate data.  
Finally, a validation routine based on partitioning first degree relatives provides a novel, rigorous standard of reproducibility.   More specifically, comparing principal angles between subspaces, in genetically related data sets, to a random direction bound carefully quantifies reproducibility. 

Even so, it is also worth considering the ways in which analyzing the HCP is distinctly challenging.  Subsetting our data along the intersection of various data blocks and first-order relatives reduces sample size.  We speculate that a larger sample size could uncover four-way partially shared structure or five-way fully shared structure whose signal was too weak to be detected in the current dataset.  
We can only say that such a space was not distinguishable from random noise within the data we analyzed.  Secondly, future work can be done in extending this analysis to other neuroscience data sets such as the Adolescent Brain Cognitive Development (\cite{CASEY2018}) (ABCD) study.  %However, the substance-use data block will be only an imperfect analogy in such a case, since the ABCD reports self-reported \textit{perceptions} of substance-use rather than use itself in the HCP. 

%perhaps discuss batch effects as well

That said, DIVAS and Jackstraw are each able to provide statistical guarantees in the HCP data despite these challenges.  As a result of the inferential structure of DIVAS, we can be confident that the subspaces that are segmented are distinct from both pure noise and other types of segmented shared, partially shared or individual space.  Moreover, Jackstraw ensures that the interpretations we draw from these subspaces are predicated upon statistically significant features not arbitrary or spurious artifacts in the data.  In short, the subspaces segmented are reproducible, interpretable, and both biologically and statistically significant.  

\begin{table}[H]
\caption{Comprehensive Principal Angle Analysis Across Original and Validation Run}\label{paatable}
\label{table4}
\begin{tabular*}{\textwidth}{@{\extracolsep\fill}cccc}
\toprule
%& \multicolumn{3}{c}{Genetics} \\
\cmidrule{1-4}
Space & Loadings & \textit{Min} PA & Fraction of PA Below RDB\\
\midrule
SC-Gene & Gene & $41.31^{\circ}$ & 1/1 \\
SC-Gene & SC & $56.33^{\circ}$ & 1/1 \\
SC-Cog & SC & $61.00^{\circ}$ & 3/3 \\
SC-Cog & Cog & $33.22^{\circ}$ & 2/3 \\
FC & FC & $3.78^{\circ}$ & 55/57 \\
SC & SC & $21.74^{\circ}$ & 25/28 \\
FC-Cog & FC & $21.26^{\circ}$ & 5/6 \\
FC-Cog & Cog & $9.94^{\circ}$ & 5/6 \\
FC-SC & FC & $9.35^{\circ}$ & 22/27 \\
FC-SC & SC & $9.41^{\circ}$ & 22/27 \\
Gene & Gene & $2.96^{\circ}$ & 68/96 \\
FC-Use & Use & $12.10^{\circ}$ & 3/4 \\
FC-Use & FC & $43.31^{\circ}$ & 2/4 \\
SC-Use & SC & $76.02^{\circ}$ & 1/2 \\
SC-Use & Use & $37.86^{\circ}$ & 1/2 \\
FC-Gene & FC & $30.06^{\circ}$ & 1/1 \\
FC-Gene & Gene & $66.49^{\circ}$ & 0/1 \\
Cog-Use & Cog & $65.11^{\circ}$ & 0/1 \\
Cog-Use & Use & $65.56^{\circ}$ & 0/1 \\

%\midrule
%Gene (Gene) & 8.2068\% & 14/15 \\

\bottomrule
\end{tabular*}
\end{table}

\addtolength{\textheight}{.5in}%

%\bigskip
%\begin{center}
%{\large\bf SUPPLEMENTARY MATERIAL}
%\end{center}

\bibliographystyle{apalike}
\bibliography{dissRef}

\appendix
\section{DIVAS Diagnostics and Random Direction Bound}
\label{sec:app}

We present the full DIVAS diagnostics corresponding to the original and validations runs discussed in Section \ref{data}.  Loadings diagnostics will be presented on the left and scores diagnostics on the right.  

\begin{figure}[h]
%\begin{center}
\includegraphics[width=7.5in]{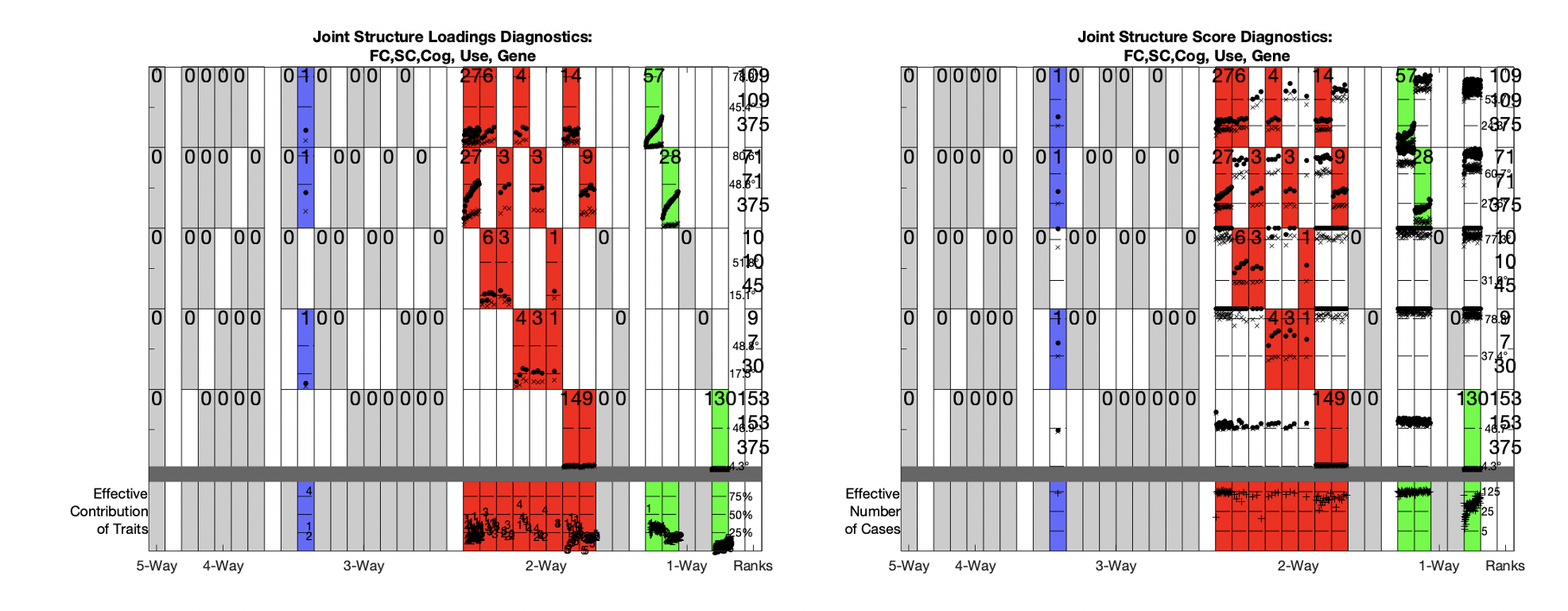}
%\end{center}
\caption{DIVAS loadings (left) and scores (right) diagnostic plot corresponding to original run.}\label{fig:diag_full}
\end{figure}

Figure \ref{fig:diag_full} shows the diagnostics corresponding to the original 375 subject HCP run.  Similar to Figure \ref{fig:diag}, each row represents a data modality, and each column represents a type of segmented shared space.  These plots distinctly offer increased rank information, angle diagnostics, and outlier assessments.  More specifically, closer examination of the far right column of each subplot will reveal that there are three ranks presented.  The second of these corresponds to the \textit{filtered rank} which is the dimension of the estimated signal subspace for that data block and was the rank reported in Figure \ref{fig:diag}.  The first and third ranks are the so-called \textit{final rank} and \textit{maximum rank} respectively.  The final rank describes the dimension of the subspace spanned by all structure (shared, partially shared, and individual) involving that block.  It is often consistent with the filtered rank, though on occasion, the final rank can be larger than the filtered rank (as is the case for substance-use in Figure \ref{fig:diag_full}).  The maximum rank is the largest possible dimension spanned by structure involving that data block, i.e. $p_k \wedge n$.  

These diagnostics also give more detail on the angle bounds, in both object and trait space, used to segment these spaces.  Within each pane, the dashed line represents the \textit{perturbation angle bound} and the dot-dashed line represents the \textit{random direction angle bound}, each described in detail in \cite{divas}.  Relative to these bounds, each direction in a particular subspace is represented by two points: $\times$ and $\bullet$.  Any $\times$ below the dashed perturbation angle bound is strong evidence that the direction can't be ruled out as joint structure for that data block.  Likewise, a $\bullet$ above the dot-dashed random direction bound indicates strong evidence that the direction can't be ruled out as an arbitrarily chosen direction with respect to that data block.  

Finally, the last row of each subplot contains information on drivers of variability and potential outliers.  That is to say, the left subplot of Figure \ref{fig:diag_full} reports the \textit{effective contribution of traits} used in segmenting a direction within a particular subspace.  Similarly, the right subplot lists the \textit{effective number of cases} used in segmenting a direction within a particular subspace, plotted on logarithmic scale.  Any direction with a particularly small effective number of cases, indicates that this direction may be driven by an outlying data object.  Similarly, any direction with a small effective contribution of traits indicates that the corresponding loadings should be driven by very few features.  

%decrease the font size in diagnostics.  

\begin{figure}[h]
\begin{center}
\includegraphics[width=7.5in]{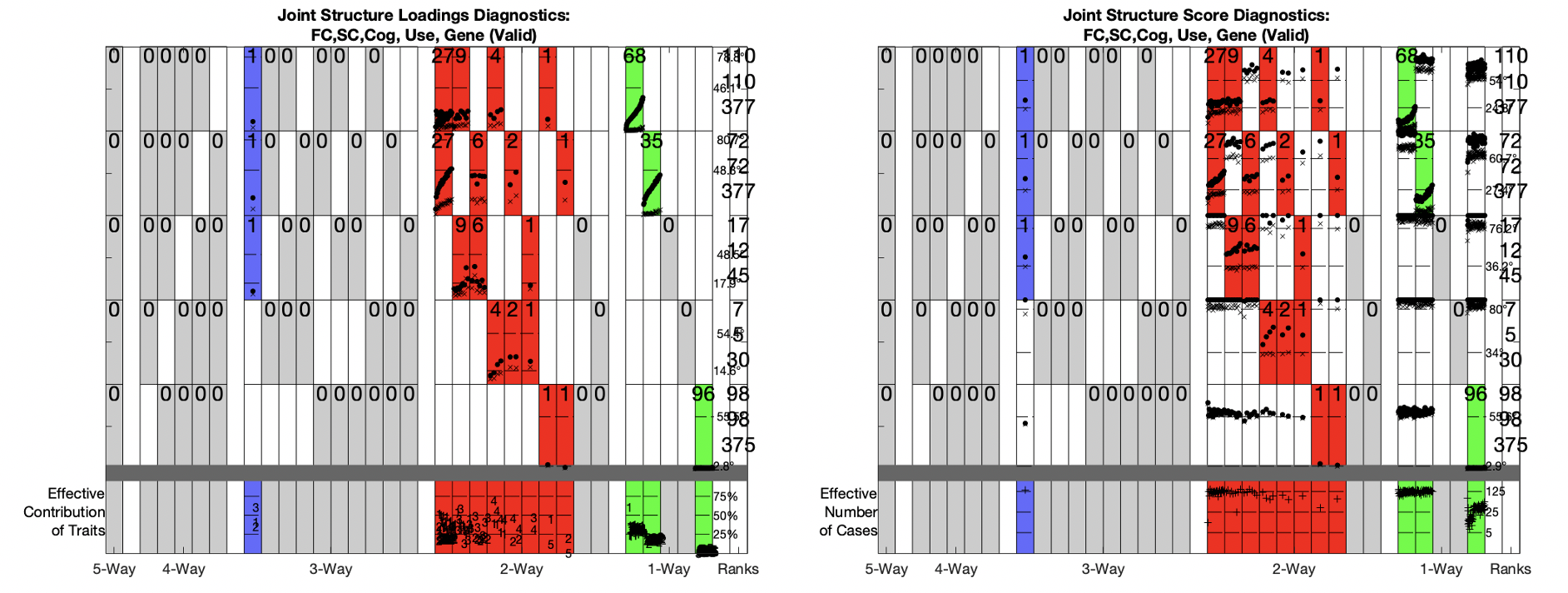}
\end{center}
\caption{DIVAS loadings (left) and scores (right) diagnostic plot corresponding to validation run.}\label{fig:valid_full}
\end{figure}

Figure \ref{fig:valid_full} presents the analogous full diagnostic plots for the 377 subject HCP validation run.  These diagnostics can be read in exactly the manner described above, so we will only take time to linger over the significance of the random direction bound.  Again, this is the dot-dashed line near the top of each pane in both loadings and scores space.  For example, the random direction bound, in loadings space, corresponding to the FC data block in the validation run is $78.8^\circ$. 

Recall that the random direction bound played a crucial role in assessing the reproducibility of subspaces within our principal angle analysis (Section \ref{val}).  More carefully, the random direction bound used in column four of Table \ref{table4}, is the minimum loading space random direction bound between original and validation runs.  As an example for SC, the loadings space random direction bounds are $80.6^\circ$ (original) and $80.7^\circ$ (validation), so our random direction bound threshold becomes $80.6^\circ$.  Any principal angle in an SC loading surpassing $80.6^\circ$, does not contribute positively to the fraction of principal angles surpassing random direction bound.  We choose the minimum of the two random direction bounds to give a conservative estimate of reproducibility.

\end{document}